\documentclass[reprint,
superscriptaddress,
nofootinbib,
amsmath,amssymb,
aps,
pra,
]{revtex4-2}
\usepackage{natbib}
\usepackage{graphicx}% Include figure files
\usepackage{dcolumn}% Align table columns on decimal point
\usepackage{bm}% bold math
\usepackage{dsfont}
\usepackage{amsmath}
\usepackage{physics}
\usepackage{color}
\usepackage{soul}
\usepackage{url}
\usepackage[hidelinks]{hyperref}
\usepackage{scalerel}

\newcommand*{\paral}{\stretchrel*{\parallel}{\perp}}
\begin{document} 

\title{ Tilted Material in an Optical Cavity:\\
Light-Matter Moir\'{e} Effect and Coherent Frequency Conversion }

\author{Arshath Manjalingal}
\altaffiliation{equal contribution}
\affiliation{Department of Chemistry, Texas A\&M University, College Station, Texas 77843, USA}

\author{Saeed Rahmanian Koshkaki}%
\altaffiliation{equal contribution}
\email{rahmanian@tamu.edu}
\affiliation{Department of Chemistry, Texas A\&M University, College Station, Texas 77843, USA}

\author{Logan Blackham}
\affiliation{Department of Chemistry, Texas A\&M University, College Station, Texas 77843, USA}

\author{Arkajit Mandal}%
\email{mandal@tamu.edu}
\affiliation{Department of Chemistry, Texas A\&M University, College Station, Texas 77843, USA}

\begin{abstract}
{\footnotesize
Exciton-polaritons formed inside optical cavities offer a highly tunable platform for exploring novel quantum phenomena. Here, we introduce and theoretically characterize a light–matter moiré effect (LMME) that arises when a 2D material is tilted inside a planar optical cavity, in contrast to stacking multiple layers at a twist angle as is done in forming 2D moiré hetero-structures. We show that this geometric tilt produces emergent periodicity in the light–matter coupling, yielding displaced replicas of the polariton dispersion and flat bands near the Brillouin-zone center. Through time-dependent quantum dynamical simulations, we demonstrate that LMME enables coherent frequency conversion and remains robust against phonon-induced decoherence.  Our findings establish LMME as a new platform for engineering polariton band structures, the generation of flat bands and performing coherent frequency conversion relevant for developing polariton-based quantum devices.}
\end{abstract}
\maketitle
{\footnotesize
\section{Introduction}

Strongly coupled light-matter systems support the formation of polaritons, hybrid light–matter quasiparticles, that have emerged as a versatile platform for inducing exotic physical phenomena in a highly tunable manner~\cite{ LiangNP2024, MandalCR2023, XuNC2023, SanvittoNM2016, SandikNM2025, XiangCR2024, KockumNRP2019, SunNC2024, NagarajanJACS2021, LiangNP2024, RibeiroCS2018, BasovNp2021, LiARPC2022, ji2025selective}.  Recent experiments demonstrate that light-matter hybrid systems can exhibit Bose–Einstein condensation at room temperature~\cite{KavokinNRP2022, KeelingARPC2020, DaskalakisNM2014, GeorgakilasNC2025, AlnatahACSP2025}, polaritonic spin-Hall effect~\cite{LiangNP2024, SeptembrePRL2024, SpencerSA2021, shi2025coherentOpticalSpinHall, lekenta2018tunableLightSpinHall}, coherent ballistic propagation at room temperature~\cite{SandikNM2025, XuNC2023, Krupp2025NC, ying2025microscopic, Blackham2025Arxiv}, and unidirectional coherence transfer~\cite{YangPNAS2023}, which are key ingredients for developing next-generation quantum devices. A key open question is how to harness the long-lived light–matter coherence of polaritons as a resource for quantum information science, as the interplay between spatially varying cavity fields and various material geometries remains a largely unexplored frontier for unlocking new functionalities. 

%with the largely unexplored interplay between spatially varying cavity fields and material geometry representing a powerful new frontier for unlocking transformative functionalities.”

%In this work, we discover moiré physics inside an optical cavity behaves distinctly different than.
In this work, we demonstrate a moiré-like effect inside an optical cavity arises when tilting a material. Typically, the moiré effect arises when two periodic structures are overlaid at an angle, producing a new emergent periodicity~\cite{CaoN2018, AndreiNRM2021, kim2013NM}, as illustrated in Fig.~\ref{fig1}a for twisted graphene bilayers. As schematically illustrated in Fig.~\ref{fig1}c, the relative rotation of the Brillouin zones leads to an intersection of the Dirac cones of the two graphene layers, leading to the formation of new bands~\cite{CaoN2018, kim2013NM}. In this work, we find that a moiré‑like effect can be achieved without multilayer stacking, but instead by tilting a single-layer material within the cavity (illustrated in Fig.~\ref{fig1}b), where the light–matter coupling generates the emergent periodic structure. We refer to this as the light-matter moiré effect (LMME). In contrast to twisted graphene, the band structure modulation in LMME is structurally different. We show
that under a material tilt, replicas of the polariton band that are displaced in the reciprocal space appear (illustrated in Fig.~\ref{fig1}d) while the original polariton band remains present. Importantly, we also find that anisotropic flat bands are formed near the Brillouin-zone center. Finally, using quantum dynamics simulations, we demonstrate that LMME can be utilized to perform coherent frequency conversion and that it is robust to phonon-induced decoherence. 

It is worth noting that LMME, introduced in this work, is fundamentally different than the recently observed polaritonic spin-Hall effect, which arises due to in-plane material (refractive-index) anisotropy in filled cavities~\cite{LiangNP2024, Xiang2024JCP}. While the band replicas formed due to LMME are visually similar to those formed in the polaritonic spin-Hall effect, the polaritonic spin-Hall effect do not feature the original (unshifted) polariton bands unlike in LMME. In addition to this, these displaced polariton replicas do not display circular spin polarization unlike in the spin-Hall effect. Additionally, we find that the LMME emerges only for material thicknesses much less than the cavity thickness (distance between the cavity mirrors) and that it disappears for filled cavities. Importantly, LMME allows for coherent frequency conversion, not exhibited in the polariton spin-Hall effect, with the difference in input and output photon frequencies set by the angle of tilt for the material.

\begin{figure}
\centering
\includegraphics[width=1.0\linewidth]{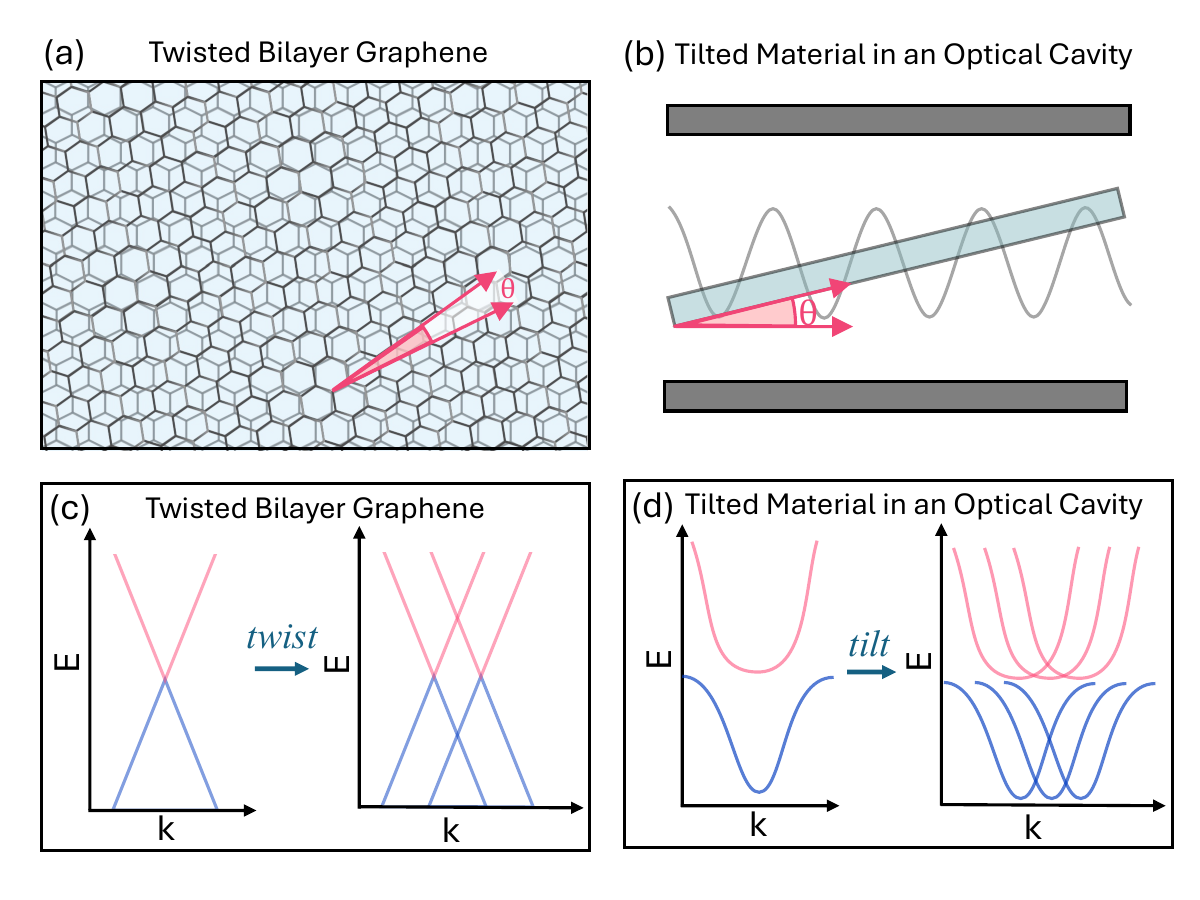}
\caption{\footnotesize \textbf{Comparison of twisted graphene and tilted material in an optical cavity} ({a}) Illustration of a graphene bilayer offset by a twist angle. ({b}) Illustration of a tilted material within a Fabry-P\'{e}rot cavity coupling to cavity radiation. ({c}) Schematic band structure of a graphene bilayer without (left) and with (right) a twist of $\theta$. ({d}) Schematic band structure of a tilted material without (left) and with (right) a tilt of $\theta$.}
\label{fig1}
\end{figure}

%%%%%%%%%%%% THEORY %%%%%%%%%%%%%%%%%
\section{Theory}
{\bf  Model.} In this work, we consider a 3D setup with 2D tilted material placed inside an optical cavity as schematically illustrated in Fig.~\ref{fig2}a and Fig.~\ref{fig2}e.  For this system we considered a multimode Holstein-Tavis-Cummings Hamiltonian~\cite{KeelingARPC2020, MandalCR2023, MandalNL2023,  sun2025exploring, ghosh2025mean}, which describes an exciton-polariton system beyond the long-wavelength approximation, and is expressed as
\begin{align}\label{RealSPaceHamiltonian}
\hat{H}_\mathrm{LM} = \hat H_\text{ex}  + \hat H_\text{cav} + \hat H_\text{int}\, ,
\end{align}
where $\hat H_\text{ex}$ and $\hat H_\text{cav}$
are the bare excitonic and cavity Hamiltonians, with $\hat H_\text{int}$ describing the exciton-cavity interactions. In this work, we consider the 2D planar material to be multilayered, with each unit cell featuring two degenerate excited states similar to a particle in a 2D box. The bare excitonic Hamiltonian is written as
\begin{equation}
    \hat H_{\text{ex}} = \sum_{\textbf{k}_{\paral},n_z} (\hat{X}^{\dagger}_{\textbf{k}_{\paral},  n_z} \hat{X}_{\textbf{k}_{\paral}, n_z} + \hat{Y}^{\dagger}_{\textbf{k}_{\paral},  n_z} \hat{Y}_{\textbf{k}_{\paral},  n_z} ) \epsilon_{ \textbf{k}_{\paral}},
\end{equation}
where ${\hat X}^{\dagger}_{\textbf{k}_{\paral},  n_z}$  and ${\hat Y}^{\dagger}_{{\bf k}_{\paral},  n_z}$ create excitons of in-plane wavevector ${\bf k}_{\paral} = k_x {\vec{x}} + k_y {\vec{ y}}$ in the $n_z$th layer. Note that,  $X^{\dagger}_{\textbf{k}_{\paral},  n_z} = \sum_{n_x,n_y} \frac{e^{-i \textbf{k}_{\paral}\cdot\textbf{R}_\textbf{n}}}{\sqrt{N_x N_y}} \hat{X}^{\dagger}_\textbf{n}$ and $Y^{\dagger}_{\textbf{k}_{\paral},  n_z}  = \sum_{n_x,n_y} \frac{e^{-i \textbf{k}_{\paral}\cdot\textbf{R}_\textbf{n}}}{\sqrt{N_x N_y}} \hat{Y}^{\dagger}_\textbf{n}$ with  $\hat{X}^{\dagger}_\textbf{n}$ and $\hat{Y}^{\dagger}_\textbf{n}$ as the exciton creation operator at the lattice site $\textbf{n} \equiv(n_x, n_y, n_z)$ located at $\textbf{R}_\textbf{n} = n_x a_x \vec{x} + n_y a_y \vec{y} + z_{\textbf{n}}\vec{z}$  with $n_\alpha \in \{0, 1, 2, \ldots, N_\alpha-1\}$  for $\alpha \in \{x,y,z\}$. Further, we consider a small angle of tilt $\theta$ in the $x$ axis, such that $ z_{\textbf{n}} = (n_x - \frac{N_x}{2})b_x + (n_z a_z + \frac{L_z}{2})$ with $b_x = a_x \sin(\theta)$. To ensure that the material is placed inside the optical cavity we impose the condition $z_{\mathbf{n}} \equiv z_{\mathbf{n}} \pmod{L_z}$. We impose a periodic boundary condition along $\vec{x}$ and $\vec{y}$ direction with the in-plane box lengths $L_x = L_y = N_x a_x = N_y a_y$. In this work, we set $a_x = a_y = 12 ~\mathrm{\AA}$, $a_z = 30 ~\mathrm{\AA}$, $N_x = N_y = 8001$ and $L_z = 5000 ~\mathrm{\AA}$.

The cavity hamiltonian $\hat{H}_\text{cav}$ is written as
\begin{equation}
    \hat{H}_{\text{cav}} = \sum_{\textbf{k}} \bigg(\hat{a}^{\dagger}_\textbf{k} \hat{a}_\textbf{k} + \hat{b}^{\dagger}_\textbf{k} \hat{b}_\textbf{k} \bigg)\omega_\textbf{k},
\end{equation}
 where
$\hat{a}^{\dagger}_\textbf{k}$ and $\hat{b}^{\dagger}_\textbf{k}$ creates a photon of polarization $s$ (TE mode) and $p$ (TM mode), respectively. Here the photonic wavevector is ${\bf k}= k_x {\vec{x}} + k_y {\vec{ y}} + k_z {\vec{ z}}$ with component along the quantization direction written as $k_z = \frac{m_z \pi}{L_z}$ where $m_z \in \{1, 2, ...\}$. In the following work, we only consider the fifth cavity mode with $m_z = 5$, although our results are valid for arbitrary $m_z$. Consequently, we denote the cavity modes only using its in-plane wavevectors, i.e. $\hat{a}_{\bf k} \rightarrow \hat{a}_{\bf k_{\paral}}$ and $\hat{b}_{\bf k} \rightarrow \hat{b}_{\bf k_{\paral}}$ as their $k_z$ component is fixed. Further, with the periodic boundary condition along $\vec{x}$ and $\vec{y}$ we write $k_x = \frac{2\pi n_x}{N_x  a_x}$ and $k_y = \frac{2\pi n_y}{N_y  a_y}$.  The corresponding photon frequency is $\omega_{\bf k} = c |\textbf{k}|/\eta$, where $\eta =2.4$  is the refractive index and $c$ is the speed of light.

The exciton-cavity interactions Hamiltonian $(\hat{H}_{\text{int}})$ is written as

\begin{align}\label{GeneralHinit}
     \hat{H}_{\text{int}} &= \sum_{\textbf{n},\textbf{k}_{\paral},j }   g\sqrt{\frac{\omega_\textbf{k}}{N}} \Bigg[  \hat{\bm{\mu}}_{\bf n}^j\cdot \left(\textbf{E}_\textbf{k}^{s}( \textbf{R}_\textbf{n})\hat{a}_{\textbf{k}_{\paral}}    + \textbf{E}_\textbf{k}^{p}( \textbf{R}_\textbf{n})\hat{b}_{\textbf{k}_{\paral}} + {\mathrm{h.c.}}\right)    \Bigg],
\end{align}
  Here, $N = N_xN_yN_z$ is total number of sites and $g$  is the light–matter coupling strength that couples the $s$ and $p$ photonic modes to the material dipoles $\bm{\mu}_{\bf n}^j \in \{\hat{\bm \mu}^x_{\textbf{n}} , \hat{\bm \mu}^y_{\textbf{n}} \}$  where $\hat{\bm \mu}^x_{\textbf{n}}  = \mu_x \left( \hat{X}_{\textbf{n}} + \hat{X}_{\textbf{n}}^\dagger \right) {\vec{x}}$ and $\hat{\bm \mu}^y_{\textbf{n}} = \mu_y \left( \hat{Y}_{\textbf{n}} + \hat{Y}_{\textbf{n}}^\dagger \right) {\vec{y}}$.  Here we consider an isotropic material and set $\mu_x = \mu_y = \mu$ and ignore the negligible dipole component in the $\vec{z}$- direction, as its contribution to the light-matter couplings is vanishingly small. The spatial variation of the radiation, $\textbf{E}_\textbf{k}^{s}( \textbf{R}_\textbf{n})$ and $\textbf{E}_\textbf{k}^{p}( \textbf{R}_\textbf{n})$,  associated to the  $s$ and $p$ photon modes, respectively, is written as~\cite{sun2025exploring, zoubi2005PRB}
 
\begin{align}
\textbf{E}_\textbf{k}^{s}( \textbf{R}_\textbf{n}) =& \sin(k_z z_{\textbf{n}})   \{\vec{e}_{\paral} \times \vec{z}\} e^{i \mathbf{k}_{\paral} \cdot \textbf{R}_{\textbf{n}}} \\
\textbf{E}_\textbf{k}^{p}( \textbf{R}_\textbf{n}) =& \bigg[ \frac{c|\textbf{k}_{\paral}|}{\omega_\textbf{k}} \cos(k_z z_{n}) \vec{z}  -i \frac{c k_z}{\omega_\textbf{k}} \sin(k_z z_{n}) \vec{e}_{\paral} \bigg] e^{i \textbf{k}_{\paral} \cdot \textbf{R}_{\textbf{n}}}, 
\end{align}

where $\textbf{e}_{\paral} = \frac{\textbf{k}_{\paral}}{|\textbf{k}_{\paral}|}$ is the unit vector  along $\textbf{k}_{\paral}$. Following a set of simplifying approximations, which includes the rotating wave approximation and then using the partially Fourier transformed exciton operators $\hat{X}^{\dagger}_{\textbf{m},k_y} = \sum_{n_y} \frac{e^{-i k_y  {\bf R}_{\bf n} \cdot {\vec{y}}}}{\sqrt{N_y}} \hat{X}^{\dagger}_{\textbf{n}}$ and $\hat{Y}^{\dagger}_{\textbf{m},k_y} =  \sum_{n_y} \frac{e^{-i k_y  {\bf R}_{\bf n} \cdot {\vec{y}}}}{\sqrt{N_y}}  \hat{Y}^{\dagger}_{\textbf{n}}$ where $ \textbf{m} \equiv (n_x,n_z)$  we obtain the following expression for the light-matter interaction Hamiltonian with further details in the Supporting Information (SI), 
\begin{align}{\label{H_interaction}}
\hat{H}_{\text{int}} &= \Omega_0 \sum_{\textbf{m},k_y} \bigg(\hat{X}^{\dagger}_{\textbf{m},k_y} \hat{A}_{n_x,{ k}_y} +  \hat{Y}^{\dagger}_{\textbf{m},k_y} \hat{B}_{n_x,k_y}   + {\mathrm{h.c.}}\bigg)\sin{(k_zz_\textbf{m})}
\end{align}
where $\Omega_0 = \frac{\sqrt{\omega_0}\mu g}{\sqrt{N_z}}$ (we set $\sqrt{\omega_0}\mu g = 0.35$ eV), $z_\textbf{m} = z_{n_x, n_z}  = (n_x - \frac{N_x}{2})b_x + (n_z a_z + \frac{L_z}{2})$ and we introduce the elliptically polarized photonic operators $\hat{A}_{\textbf{k}}$ and $\hat{B}_{\bf k}$ which are defined as 

\begin{align}
     \hat{A}_{n_x, k_y}=\sum_{k_x}\frac{e^{ik_xn_xa_x}}{\sqrt{N_x }} \hat{A}_{\textbf{k}_{\paral}}=\sum_{k_x} \frac{e^{ik_xn_xa_x}}{\sqrt{N_x \mathcal{S}_{x}({\bf k})}}  \bigg(i k_y \hat{a}_{\textbf{k}_{\paral}} - \frac{ck_z}{\omega_k} k_x\hat{b}_{\textbf{k}_{\paral}}\bigg)  \nonumber \\
    \hat{B}_{n_x, k_y}=\sum_{k_x}\frac{e^{ik_xn_xa_x}}{\sqrt{N_x }} \hat{B}_{\textbf{k}_{\paral}}=\sum_{k_x} \frac{e^{ik_xn_xa_x}}{\sqrt{N_x \mathcal{S}_{y}({\bf k})}}  \bigg(i k_x \hat{a}_{\textbf{k}_{\paral}} + \frac{ck_z}{\omega_k} k_y\hat{b}_{\textbf{k}_{\paral}}\bigg) 
 \end{align}
 where $\mathcal{S}_{x} = {{1 - \frac{c^2 k_x^2}{\omega_{\bf k}^2}}}$ and $\mathcal{S}_{y} = {{1 - \frac{c^2 k_y^2}{\omega_{\bf k}^2}}}$. This transformation allow us to write the full Hamiltonian into two non-interacting parts as 

\begin{align}
    \hat{{H}}_\mathrm{LM} = \hat{\mathcal{H}}_\mathrm{AX}(\{\hat{X}_{\textbf{k}_{\paral}, n_z}, \hat{A}_{\textbf{k}_{\paral}}\}) + \hat{\mathcal{H}}_\mathrm{BY}(\{\hat{Y}_{\textbf{k}_{\paral}, n_z}, \hat{B}_{\textbf{k}_{\paral}}\}),
\end{align}
where $\hat{\mathcal{H}}_\mathrm{AX}(\{\hat{X}_{\textbf{k}_{\paral}, n_z}, \hat{A}_{\textbf{k}_{\paral}}\})$ and $\hat{\mathcal{H}}_\mathrm{BY}(\{\hat{Y}_{\textbf{k}_{\paral}, n_z}, \hat{B}_{\textbf{k}_{\paral}}\})$ have identical structure.   Due to this, we study the dynamics of our light-matter Hamiltonian by focusing only on $\hat{\mathcal{H}}_\mathrm{AX}(\{\hat{X}_{\textbf{k}_{\paral}, n_z}, \hat{A}_{\textbf{k}_{\paral}}\})$ which is expressed as
\begin{align}\label{real-space-LM}
    \mathcal{\hat{H}}_\mathrm{AX} &= \sum_{k_y} \mathcal{\hat{H}}_\mathrm{AX}^{k_y} \\
    &= \sum_{k_y} \bigg[ \sum_{k_x} \Big( \hat{A}^\dagger_{\mathbf{k}_{\paral}} \hat{A}_{\mathbf{k}_{\paral}} \omega_{\mathbf{k}} + \sum_{n_z}\hat{X}_{\mathbf{k}_{\paral},n_z}^{\dagger} \hat{X}_{\mathbf{k}_{\paral},n_z} \epsilon_{\mathbf{k}_{\paral}} \Big) \nonumber \\
    &+  {\Omega_0}\sum_{\mathbf{m}} \sin{(k_zz_\textbf{m})}\Big( \hat{X}^{\dagger}_{\mathbf{m},k_y} \hat{A}_{n_x,k_y} + \hat{X}_{\mathbf{m},k_y} \hat{A}^{\dagger}_{n_x,k_y} \Big) \bigg]\nonumber
\end{align}
This formulation enables us to perform independent 2D simulations at each $ k_y $, which can be combined to reconstruct the full 3D dynamics of the system. This Hamiltonian also illustrates that the periodicity of the system can be modulated by the angle of tilt $\theta$, as it modifies the spatially oscillatory light–matter coupling term $\sin(k_z z_\mathbf{m})$, where $z_\mathbf{m}$ depends on $\theta$. This is the origin of the LMME studied in this work. To gain intuition into the band structure modification of the polariton for tilted systems, we perform a Fourier transform along the $\vec{x}$- direction. This allows us to rewrite the Hamiltonian  $\mathcal{\hat{H}}_\mathrm{AX}$ in the reciprocal space as,
\begin{align}\label{k-space-LM}
        &\mathcal{\hat{H}}_\mathrm{AX} = \sum_{{\bf k}_{\paral}}  \Big( \hat{A}^\dagger_{\mathbf{k}_{\paral}} \hat{A}_{\mathbf{k}_{\paral}} \omega_{\mathbf{k}} + \sum_{n_z}\hat{X}^{\dagger}_{\mathbf{k}_{\paral},n_z} \hat{X}_{\mathbf{k}_{\paral},n_z} \epsilon_{\mathbf{k}_{\paral}} \Big)   \\
    &+\frac{\Omega_0}{2i} \sum_{n_z,{\bf k}_{\paral}} \hat{X}^{\dagger}_{\mathbf{k}_{\paral}, n_z}\Big( \hat{A}_{{\bf k}_{\paral} + {\bf \Delta k_x }}e^{i\phi_z}  
    +  \hat{A}_{{\bf k}_{\paral} - {\bf \Delta k_x }}e^{-i\phi_z} \Big)+ \mathrm{h.c.} \nonumber
\end{align}

Here, ${\bf \Delta k_x } = { \Delta k_x } \vec{x} = k_z\sin(\theta) \vec{x}$ and  $\phi_z = k_z\big(\frac{-N_xb_x + L_z}{2} + n_z a_z \big)$. The light–matter Hamiltonian presented in Eq.~\ref{k-space-LM} illustrates that two photon modes differing by $\pm 2{\bf \Delta k_x }$ in the $x$ component of their wavevectors (given a tilt along $x$) {\it effectively} couple to each other through their interaction with the exciton.

 \begin{figure*}
\centering
  \includegraphics[width=1.0\linewidth]{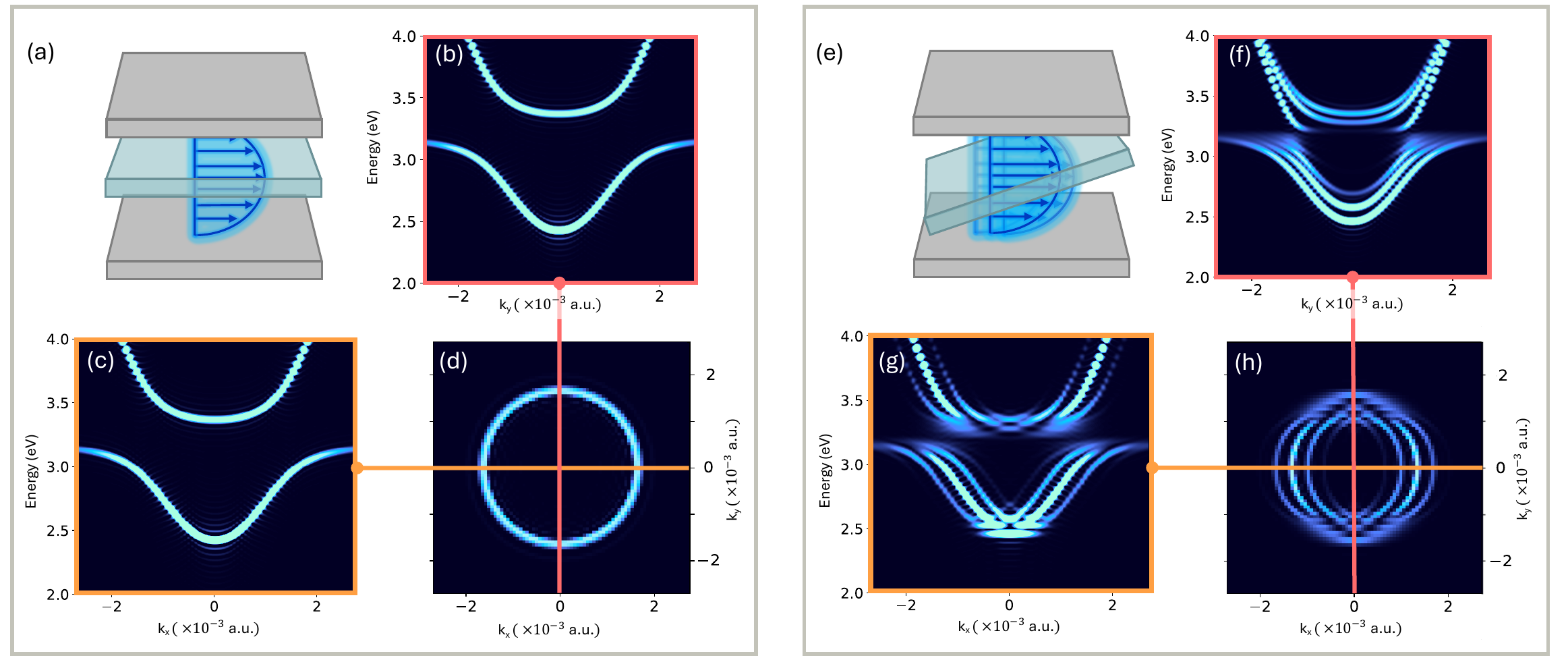} 
\caption{\footnotesize \textbf{Two-dimensional band structure of a flat and tilted material ($\theta = 5.98^\circ$).} ({a}) Schematic representation of a flat material inside an optical cavity, coupled to cavity radiation. ({b-c}) One dimensional band structures along Y and X directions, respectively for a flat material inside a cavity. ({d}) Two dimensional band structure at E = $3$ eV of a flat material inside an optical cavity with Y (red) and X (orange) directional cuts, correlating with ({b}) and ({c}) respectively. ({e}) Schematic representation of a tilted material of angle $\theta$ in the X direction, inside an optical cavity, coupled to cavity radiation. ({f-g}) One dimensional band structures along Y and X directions, respectively for a tilted material inside a cavity. ({h}) Two dimensional band structure at E = $3$ eV of a tilted material inside an optical cavity with Y (red) and X (orange) directional cuts, correlating with ({f}) and ({g}) respectively. }
\label{fig2}
\end{figure*}

\section{Results and Discussion}

Fig. 2 presents the angle-resolved polariton spectra of a 2D material within an optical cavity without a tilt (a-d) and with a tilt along the $x$-direction (e-h) computed from the photonic spectral function written as~\cite{Blackham2025Arxiv, Nguyen2025Arxiv}

\begin{align}
I(\omega, {\bf k}_{\paral}) = \Re\Bigg[\lim_{T\rightarrow \infty} \int_{0}^T dt~ e^{i \omega t} \big\langle  \bar{0}|\hat{A}_{{\bf k}_{\paral}}| {\Psi} (t) \big\rangle_ \cdot \cos({\pi t}/{2T})\Bigg], \nonumber
\end{align}
where $| {\Psi} (0)\rangle = \hat{A}_{{\bf k}_{\paral}}^{\dagger} |\bar{0}\rangle$ with $\ket{\bar{0}}$ as the vacuum state and $| {\Psi} (t)\rangle$ is the exciton-polariton wavefunction at time $t$ obtained by solving the time-dependent Schr\"{o}dinger equation (see details in the SI). 

Fig. \ref{fig2}a shows a schematic illustration of a 2D material (without tilt) placed inside a Fabry–Pérot cavity, where the cavity field is quantized along the $\vec{z}$‑direction. The polaritonic dispersions along $k_y$ (at $k_x = 0$) and along $k_x$ (at $k_y = 0$) are displayed in Fig. \ref{fig2}b-c, respectively. Both figures exhibit the two characteristic polariton bands formed by the coupling between a photonic band and an excitonic band~\cite{MandalCR2023, KeelingARPC2020}, and they (Fig. \ref{fig2}b and \ref{fig2}c) are identical due to the isotropic nature of the system. Consequently, Fig. \ref{fig2}d displays a 2D cut (at $E = 3$ eV) of the polaritonic dispersion which appears as a perfect circle as expected. 
\begin{figure*}
\centering
\includegraphics[width=1.0\linewidth]{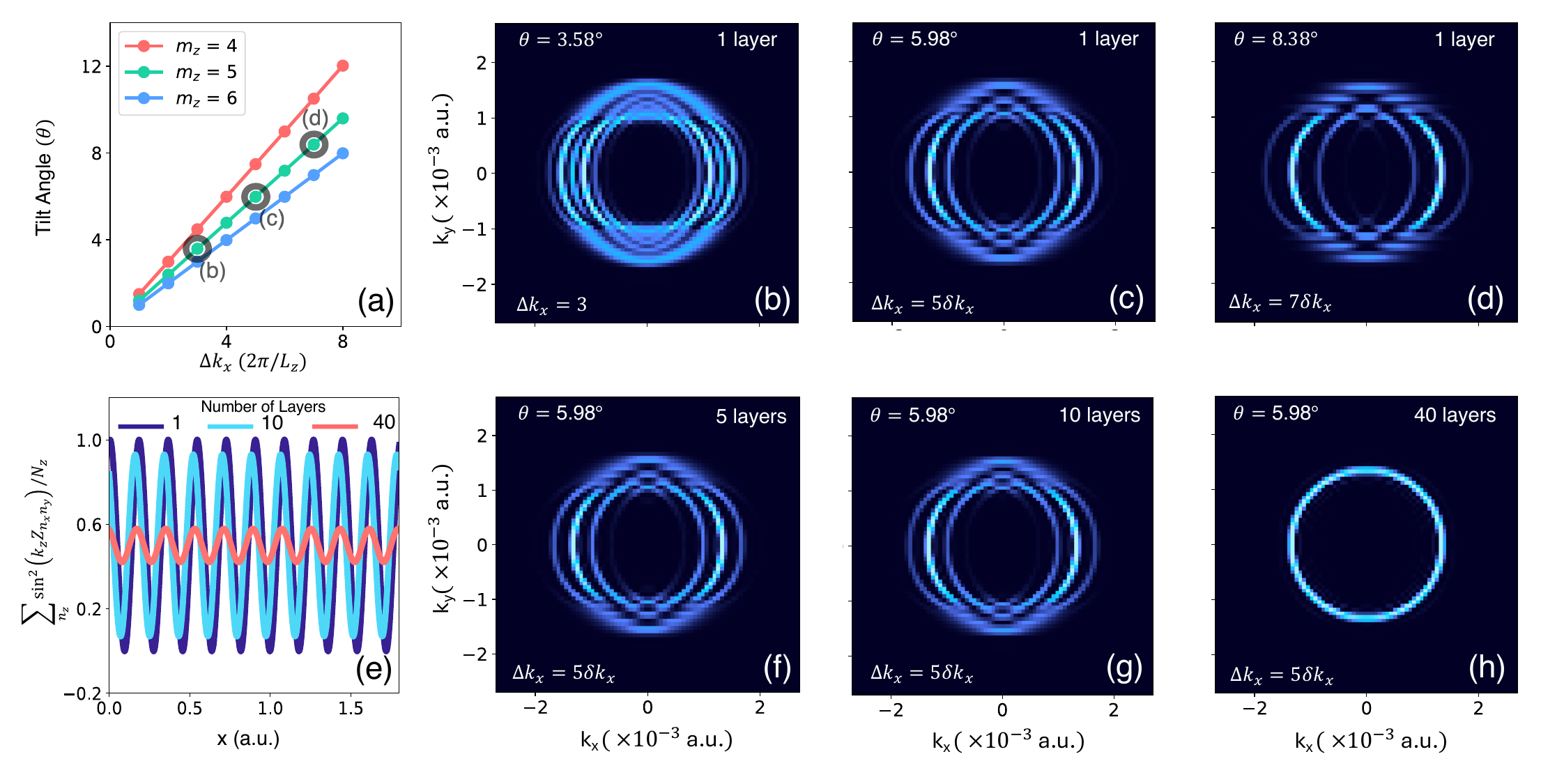}
\caption{\footnotesize \textbf{Tilt-induced separation of side bands and their dependence on  material thickness.} ({a}) Relation between tilt angle $\theta$ and $\Delta k_x$ for the model system. {(b-d)} Two dimensional band structures for tilts of $\theta = 3.58^{\circ}, 5.98^{\circ}, 8.38^{\circ}$ corresponding to $\Delta{k_x}$, where $\delta k_z = 2\pi/L_x$ and $p \in\{3,5,7\}$, respectively for a single layer. ({e}) Normalized spatially varying (bright-layer) light matter coupling at $n_z = 1, 10, 40$. ({f-h}) Two dimensional band structures of tilt $\theta = 5.98^{\circ}$ corresponding to $\Delta{k_x} = 5\delta k_x$ for $n_z = 5, 10, 40$, respectively. We use $N_x = N_y = 8001$.}
\label{fig3}
\end{figure*}
In Fig. \ref{fig2}e–h, we examine exciton-polariton bands formed by coupling a single layer material, with a small tilt ($\theta = 5.7^\circ$), to an optical cavity  (schematically illustrated in Fig. \ref{fig2}e). Fig. \ref{fig2}g shows the polariton dispersion along $k_x$ at $k_y = 0$, which features multiple replicas of the original polaritonic bands displaced along $k_x$, together with the original unshifted polaritonic bands $E_\pm (k_x, k_y) \approx \frac{1}{2}[\omega_{\bf k} + \epsilon_0 ]  \pm \frac{1}{2}\sqrt{4\Omega_0^2 + (\omega_{\bf k} - \epsilon_0 )^2 }$ assuming $\epsilon_{\bf k_{\paral}} \approx \epsilon_0$ (since we focus on ${\bf k_{\paral}} \rightarrow 0$). This polaritonic dispersion, which is the defining characteristic of the LMME, can be understood by inspecting Eq.~\ref{k-space-LM}. Eq.~\ref{k-space-LM} shows that a photonic mode $\hat{A}_{\bf k_{\paral}}$ {\it effectively} couples with the photonic modes $\hat{A}_{{\bf k_{\paral}} \pm 2\alpha{\bf \Delta k_x} }$, with $\alpha \ge 1$ and an integer, through the light-matter coupling to excitons. However, this coupling fades exponentially with increasing $\alpha$ with the effective coupling at $\alpha = 1$ being the strongest. 

%The coupling for $\alpha = 1$, i.e. the effective coupling between  $\hat{A}_{\bf k_{\paral}}$ and $\hat{A}_{\bf k_{\paral} \pm 2{\bf \Delta k_x} }$ can be expressed using second-order perturbation theory as
% \begin{align}
% \approx \frac{\Omega_0^2 e^{2i\phi}}{4(\omega_{{\bf k } } - \epsilon_0)} + \frac{\Omega_0^2 e^{2i\phi}}{4(\omega_{{\bf k }  \pm 2{\bf \Delta k}}- \epsilon_0)}.
% \end{align}

As a result of this effective coupling, for a given $k_x$ (with $k_y = 0$ fixed) in addition to the original polariton bands $E_\pm(k_x, k_y = 0)$, new bands at $E_\pm(k_x \pm { \Delta k_x }, k_y = 0)$  emerge. These bands are clearly visible in Fig.~\ref{fig2}g. Importantly, these band replicas also strongly interact with the original polariton bands near ${\bf k}_{\paral} \rightarrow {\bf 0}$ and leads to the formation of flat bands. We anticipate that this non-dispersive character of the polariton flat bands will open new opportunities for inducing exotic physical phenomena, as observed in 2D moiré heterostructures~\cite{Neves2024NPJCM, cao2018correlated, park2021flavour, balents2020superconductivity}. Specifically, the emergence of flat bands implies a substantial enhancement of the density of states at ${\bf k_{\paral}} \rightarrow {\bf 0}$, which may facilitate the formation of polariton condensates at room temperature. While this lies beyond the scope of the present work, LMME-induced polariton condensate formation will be studied in our future work.

Fig.~\ref{fig2}f shows the polariton dispersion along $k_y$ (orthogonal to the tilt direction) at $k_x = 0$, featuring multiple energy-shifted replicas of the original polaritonic bands. All polariton bands near $k_y = 0$ exhibit finite curvature, indicating a finite effective mass. This illustrates that tilting a (isotropic) material along one direction leads to an anisotropic flatband, which is expected to lead to anisotropic polariton transport and anisotropic localization. 

Fig.~\ref{fig2}h display the 2D cut of the polariton dispersion at a fixed energy $E = 3$ eV. In comparison to the untilted scenario, presented in Fig.~\ref{fig2}d, Fig.~\ref{fig2}h shows (at least) three circles with two displaced along the $x$ direction which closely resemble a Rashba–Dresselhaus-like splitting seen in polaritonic spin-Hall effect~\cite{LiangNP2024}. 
Notably, unlike the spin–Hall effect, the original polariton band does not disappear in the LMME. The displacement between the band replicas shown in Fig.~\ref{fig2}g-h depends on the angle of tilt $\theta$ and is given by ${ \Delta k_x }   = k_z\sin(\theta)  = \frac{m_z \pi}{L_z} \sin(\theta)$.  Therefore a larger displacement in reciprocal space can be achieved at a smaller tilt angle when using a higher cavity mode $m_z$.  In analogy to the Rashba–Dresselhaus Hamiltonian~\cite{LiangNP2024}, these (uncoupled) side bands can be {\it crudely} described by 
\begin{align}
    \hat{\mathcal{H}}_{\pm} = -\frac{\hbar^2}{2m_{\mathrm{eff}}}\nabla_{\paral}^2 \pm 2 i k_z\sin(\theta)\frac{\hbar}{m_{\mathrm{eff}}} \frac{\partial}{\partial x},
\end{align}
where $m_{\mathrm{eff}}$ is the effective mass extracted from the curvature of the polaritonic dispersion. We emphasize, however, that unlike in the polariton spin–Hall effect, these $\hat{\mathcal{H}}_{\pm}$ do not correspond to two circularly polarized light modes; in our case, both side bands are elliptically polarized. 

%Note that when considering the full Hamiltonian ($\hat{{H}}_\mathrm{LM} = \hat{\mathcal{H}}_\mathrm{AX} + \hat{\mathcal{H}}_\mathrm{BY}$) there are two degenerate polariton 

In Fig.~\ref{fig3}, we illustrate how varying the tilt angle and stacking multiple material layers influence the polariton dispersion and the resulting LMME. As illustrated in Fig.~\ref{fig3}a, the angle of tilt $\theta$ increases monotonically (and almost linearly) with $\Delta k_x$. This is expected as at small $\theta$ as ${ \Delta k_x }   \approx  \frac{m_z \pi}{L_z} \theta$. This monotonic increase in $\Delta k_x$ can also be seen in Fig.~\ref{fig3}b-d, which shows a 2D cut of the polariton dispersion at $\theta = 3.58^\circ, 5.98^\circ$ and $8.38^\circ$, respectively. At small $\theta$, more than two displaced polariton bands appear, alongside the original polariton band, seen in Fig.~\ref{fig3}b. This occurs because, at smaller displacements of $\Delta k_x$, the energy difference between photonic modes for $\alpha = 2$ (i.e., between $\hat{A}_{\mathbf{k}_{\parallel}}$ and $\hat{A}_{\mathbf{k}_{\parallel} \pm {\bf 4\Delta k_x}}$) is sufficiently small to enable appreciable hybridization, despite the significantly weaker effective couplings. Consequently, at slightly higher angles of tilt in Fig.~\ref{fig3}c-d only two displaced polaritons bands appear. 

Fig.~\ref{fig3}f-h illustrate how stacking multiple material layers influence the polariton dispersion under a constant tilt ($\theta = 5.98^\circ $) for 5, 10, and 40 layers, respectively. To provide an analytical understanding of how polaritonic dispersion are modified in a multilayered setup, consider the light-matter Hamiltonian rewritten using a bright layer formalism presented in our recent works~\cite{koshkaki2025exciton, MandalNL2023}. In this formalism a multilayered material is described using an effective (single) bright layer which couples to quantized radiation. The effective bright-layer Hamiltonian ($\mathcal{\hat{H}}_\mathrm{AX} \mapsto \mathcal{\hat{H}}_\mathrm{AX}^B$) is written as (with details provided in the SI) 

\begin{align}\label{bright-space-LM}
    \mathcal{\hat{H}}_\mathrm{AX}^B &= \sum_{\mathbf{k}_{\paral}}   \hat{A}^\dagger_{\mathbf{k}_{\paral}} \hat{A}_{\mathbf{k}_{\paral}} \omega_{\mathbf{k}} + \sum_{n_x, k_y} \Big(\hat{X}_{n_x, k_y,\mathrm{B}}^{\dagger} \hat{X}_{n_x, k_y,\mathrm{B}}\Big) \epsilon_{\mathbf{k}_{\paral}}   \\
    &+  {\Omega_0}\sum_{n_x, k_y} \sqrt{\mathcal{N}_{n_x}} \Big( \hat{X}^{\dagger}_{n_x,k_y, \mathrm{B}} \hat{A}_{n_x,k_y} + \hat{X}_{n_x,k_y, \mathrm{B}} \hat{A}^{\dagger}_{n_x,k_y} \Big),   \nonumber
\end{align}
where $\hat{X}_{n_x, k_y,\mathrm{B}}^{\dagger} = \frac{1}{\sqrt{\mathcal{N}_{n_x}}
} \sum_{n_z} \sin{(k_z z_{\bf m})} \hat{X}_{n_x, k_y,n_z}$ with $\mathcal{N}_{n_x} = \sum_{n_z}  \sin^2{(k_zz_{\bf m})}$  a normalization constant. Since $\mathcal{N}_{n_x}$ is approximately proportional to $N_z$,   $\sqrt{\mathcal{N}_{n_x}}\Omega_0$ is almost a constant when varying the number of layers in Fig.~\ref{fig3}f–h  ensuring a fair comparison. Note that all numerical results presented here employ a full description of the multilayer material and its coupling to the quantized radiation (as described in Eq.~\ref{real-space-LM}),  involving no additional approximations.

 Fig.~\ref{fig3}f-g presents the polariton dispersion when stacking 5 and 10 layers, respectively. These polariton dispersions are nearly identical to the single layer result in  Fig.~\ref{fig3}c with the same angle of tilt. This is because, at $N_z = 5$ or 10, the thickness  of the material remains much smaller than the distance between the mirrors (10$a_z$ = 300  \AA $~\ll L_z$) such that it can be effectively regarded as single layer material. Mathematically,   $\mathcal{N}_{n_x} \approx N_z \sin^2(k_z z_{n_x, 0})$ such that Eq.~\ref{bright-space-LM} reduces to the single layer version of the $\mathcal{\hat{H}}_\mathrm{AX}$ in Eq.~\ref{real-space-LM}, i.e. $\mathcal{\hat{H}}_\mathrm{AX}^{B} \rightarrow \mathcal{\hat{H}}_\mathrm{AX} (N_z = 1)$ for $N_z a_z \ll L_z$. This suggests that LMME is not limited to a single layer material but it can be observed as long as the material thickness is much smaller than the thickness of the optical cavity. 

\begin{figure*}
\centering
\includegraphics[width=1.0\linewidth]{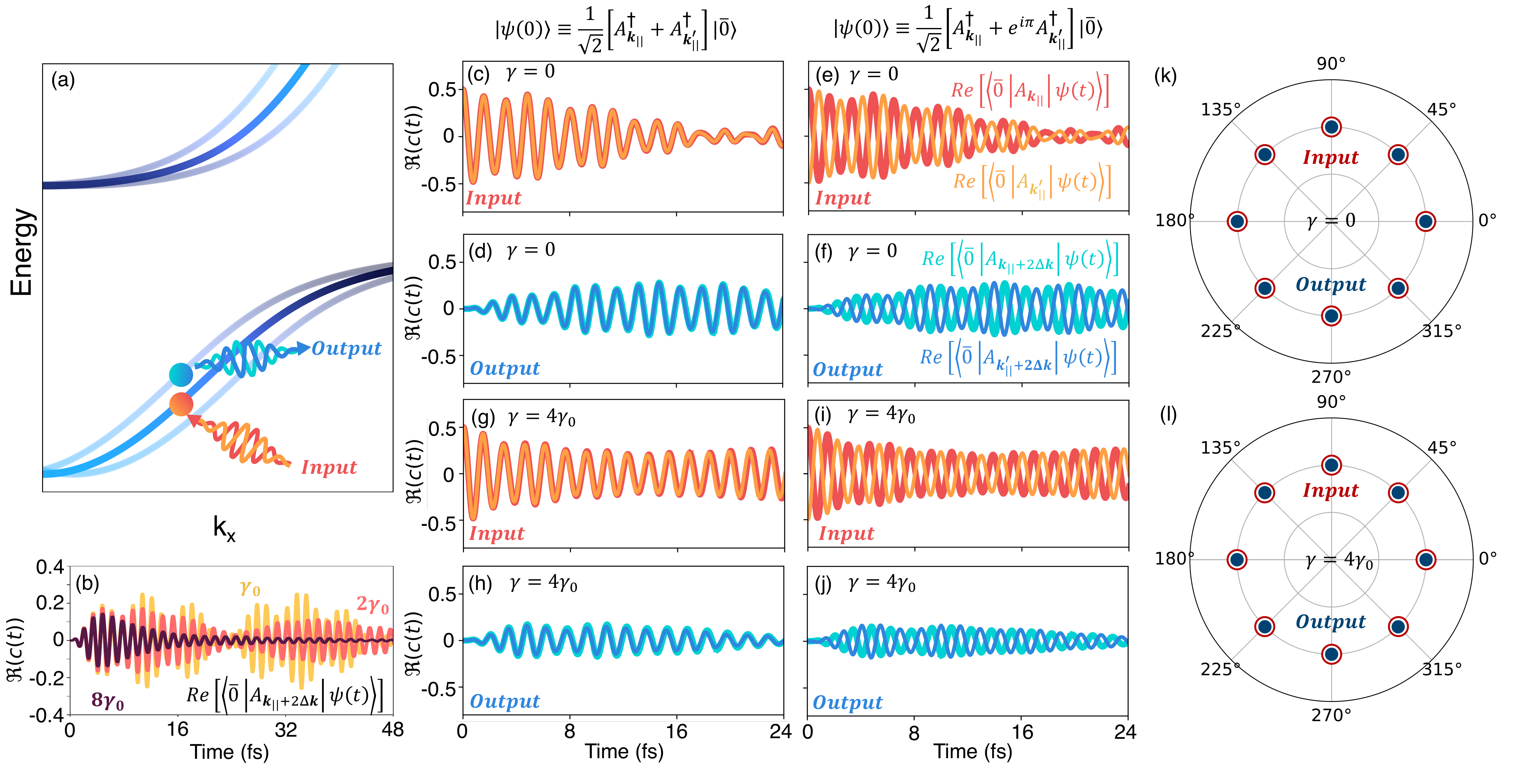}
\caption[]{\footnotesize \textbf{Coherent frequency conversion.} (a) Band structure depicting an input coherent superposition $\frac{1}{\sqrt{2}} \Big[ \hat{A}^{\dagger}_{\bf k_{\paral}}  + e^{i\phi}\hat{A}^{\dagger}_{\bf k'_{\paral}}\Big]|\bar{0}\rangle$, and an output coherent superposition of states $\frac{1}{\sqrt{2}} \Big[ \hat{A}^{\dagger}_{\bf k_{\paral} + 2\Delta k}  + e^{i\phi}\hat{A}^{\dagger}_{\bf k'_{\paral}+ 2\Delta k}\Big]|\bar{0}\rangle$, for the material at angle of tilt $\theta = 5.98^{\circ}$ corresponding to $\mathbf{\Delta k} = 5\delta k_x$. (b) Real component of state $\hat{A}^{\dagger}_{\bf k_{\paral}}|\bar{0}\rangle$ with $|\psi(t)\rangle$ over 48 fs, with phonon couplings equal to $\gamma_0, 2\gamma_0, 8\gamma_0$, where $\gamma_0 = 7.3067 \times 10^{-5}$ a.u. (c-d) The input and output superposition over 24 fs, without phonon coupling or an initial phase difference between states $\hat{A}^{\dagger}_{\bf k_{\paral}}|\bar{0}\rangle$ and $\hat{A}^{\dagger}_{\bf k'_{\paral}}|\bar{0}\rangle$. (e-f) The input and output superposition over 24 fs, without phonon coupling and an initial phase difference of $\phi = \pi$ between states $\hat{A}^{\dagger}_{\bf k_{\paral}}|\bar{0}\rangle$ and $\hat{A}^{\dagger}_{\bf k'_{\paral}}|\bar{0}\rangle$. (g-j) Mimic the structure of (c-f), differing as the phonon coupling $\gamma = 4\gamma_0$. (k-l) Polar plots displaying the initial phase difference (after $~.5$ fs) of the input and output in $45^{\circ}$ intervals, for $\gamma = 0$ (k) and $\gamma = 4\gamma_0$ (l).}
\label{fig4}
\end{figure*}

 Fig.~\ref{fig3}h presents the polaritonic dispersion when stacking 40 layers which corresponds to a material thickness of $1200$ \AA. At this thickness, the LMME fully disappears, and the polaritonic dispersion resembles that of the untilted material (see Fig.~\ref{fig2}d). This behavior can be understood by inspecting $\mathcal{N}_{n_x}/N_z$, which characterizes the spatially varying light–matter couplings under material tilt. Fig.~\ref{fig3}e plots $\mathcal{N}_{n_x}/N_z$ as a function of $x$. In the single-layer case ($N_z = 1$), $\frac{\mathcal{N}_{n_x}}{N_z} = \sin^2\bigl(k_z\,z_{n_x,0}\bigr),$
which results in strong spatial oscillations (dark blue solid line in Fig.~\ref{fig3}e) that give rise to the LMME. Importantly, for $N_z = 10$, $\mathcal{N}_{n_x}/N_z$ remains oscillatory and closely matches the single-layer case. At $N_z = 40$, the oscillation is greatly reduced and to a good approximation $\mathcal{N}_{n_x}/N_z$ could be regarded as a constant. Note that replacing $\mathcal{N}_{n_x}$ to a spatially independent constant reduces the Hamiltonian in Eq.~\ref{bright-space-LM} to the untilted scenario. This explains why the polariton dispersion in  Fig.~\ref{fig3}h resembles the untilted scenario in Fig.~\ref{fig2}d. 

Overall, LMME induced by tilting a material inside an optical cavity can unlock exotic new phenomena, including the formation of momentum-displaced polariton bands and anisotropic flat bands. Below, we illustrate one exciting application of LMME: coherent frequency conversion, which is of enormous importance in quantum information processing \cite{Mann2023qfc, wang2023qfc, curtz2010coherent, Abdo2013CFC}.

Fig.~\ref{fig4} illustrates the applicability of using LMME for performing coherent frequency conversion. This process is schematically illustrated in Fig.~\ref{fig4}a. We prepare an initial superposition of two single-photon modes with relative phase $\phi$ written as

\begin{align}
    |\Psi (0)\rangle = \frac{1}{\sqrt{2}} \Big[ \hat{A}^{\dagger}_{\bf k_{\paral}}  + e^{i\phi}\hat{A}^{\dagger}_{\bf k'_{\paral}} \Big]|\bar{0}\rangle.
\end{align}

Here we choose ${\bf k}_{\paral} = 20\frac{2\pi}{L_z}  \vec{x} + 0 \vec{y}$ and ${\bf k'}_{\paral} = 21\frac{2\pi}{L_z}    \vec{x} + 0 \vec{y}$ for preparing our initial condition which corresponds to a photon frequency of $\omega_{\bf k_{\paral}} = 2.798$ eV and $\omega_{\bf k_{\paral}'} = 2.819$ eV, respectively. Given this initial state, we detect two photon at ${\bf k}_{\paral} + 2{\bf \Delta k}$ and ${\bf k}_{\paral}' + 2{\bf \Delta k}$ (corresponds to a photon frequency of $\omega_{\bf k_{\paral}} = 3.046$ eV and $\omega_{\bf k_{\paral}'} = 3.075$ eV, respectively for a tilt angle $\theta = 5.98^\circ$ with $N_z = 1$) and measure their phase difference. Specifically, we compute $\langle  \bar{0}| \hat{A}_{\bf k_{\paral} + 2{\bf \Delta k}} |\Psi (t)\rangle$  and $\langle  \bar{0}| \hat{A}_{\bf k'_{\paral} + 2{\bf \Delta k}} |\Psi (t)\rangle$. 

% The difference in the input and output photon frequency is expressed as
% \begin{align}
% \Delta \omega (\theta) = \omega_{\bf k_{\paral} + 2{\bf \Delta k}} - \omega_{\bf k_{\paral}} \approx    \frac{2 c^2}{\eta^2 \omega_{\bf k_{\paral}}} { \Delta k}_x ({ \Delta k}_x + k_x).
% \end{align}

Fig.~\ref{fig4}c–h present numerical results demonstrating coherent frequency conversion enabled by tilting the material inside an optical cavity. In Figs.~\ref{fig4}c–d, we set the initial relative phase $\phi=0$ and simulate the dynamics in the absence of any phonons. Fig.~\ref{fig4}c plots $\langle\bar{0}|\hat{A}_{{\bf k}_{\paral}}|\Psi(t)\rangle$ (red solid line) and $\langle\bar{0}|\hat{A}_{{\bf k'_{\paral}}}|\Psi(t)\rangle$ (orange solid line) as functions of time, both of which lie on top of each other, as expected for $\phi=0$. The outputs, $\langle\bar{0}|\hat{A}_{{\bf k}_{\paral}+2\Delta{\bf k}}|\Psi(t)\rangle$ and $\langle\bar{0}|\hat{A}_{{\bf k'}_{\paral}+2\Delta{\bf k}}|\Psi(t)\rangle$ (cyan and blue solid lines in Fig.~\ref{fig4}d), also lie on top of each other, indicating that the output phase difference (at $t\rightarrow 0$) is $0^\circ$.

In Figs.~\ref{fig4}e–f, we set the initial relative phase to $\phi=\pi$ and simulate the dynamics in the absence of phonons. Consequently, Fig.~\ref{fig4}e shows that $\langle\bar{0}|\hat{A}_{{\bf k}_{\paral}}|\Psi(t)\rangle$ (red solid line) and $\langle\bar{0}|\hat{A}_{{\bf k'_{\paral}}}|\Psi(t)\rangle$ (orange solid line) oscillate out of phase by $\phi = \pi$. Importantly, Fig.~\ref{fig4}f shows that $\langle\bar{0}|\hat{A}_{{\bf k}_{\paral}+2\Delta{\bf k}}|\Psi(t)\rangle$ and $\langle\bar{0}|\hat{A}_{{\bf k'}_{\paral}+2\Delta{\bf k}}|\Psi(t)\rangle$ also oscillate out of phase, mirroring the input phase difference by $\pi$. These results illustrate that a superposition of two photons at ${\bf k}_{\paral}$ and ${\bf k}_{\paral}'$ can be converted to a superposition of two photons at  ${\bf k}_{\paral} + 2\Delta{\bf k}$ and ${\bf k}_{\paral}' + 2\Delta{\bf k}$ which carry the same relative phase encoded in the initial state. Fig.~\ref{fig4}k, shows that this scheme is broadly applicable, where we have plotted the relative initial phase difference between the input and output phase that is evaluated as $\phi_{\mathrm {output} }=\lim_{t\rightarrow 0} \arg\left[ \frac{\langle\bar{0}|\hat{A}_{{\bf k}_{\paral}+2\Delta{\bf k}}|\Psi(t)\rangle }{ \langle\bar{0}|\hat{A}_{{\bf k}_{\paral}'+2\Delta{\bf k}}|\Psi(t)\rangle} \right]$ with more numerical details provided in the SI.  

To assess the feasibility of performing coherent frequency conversion in the presence of phonon induced disorder, we simulate the following 1D generalized Holstein-Tavis-Cummings Hamiltonian written as

\begin{align}
    \hat{H}_{\mathrm{HTC}} = \mathcal{\hat{H}}_\mathrm{AX}^{k_y = 0} + \sum_{\mathbf{m}}\hat{b}_{\mathbf{m}}^{\dagger}  \hat{b}_{\mathbf{m}} \omega_{\mathrm{b}} + \frac{\gamma}{\sqrt{2\omega_{\mathrm{b}}}} \sum_{\bf m} \hat{X}^{\dagger}_{\mathbf{m},0} \hat{X}_{\mathbf{m},0}  (\hat{b}_{\mathbf{m}}^{\dagger} + \hat{b}_{\mathbf{m}}), \nonumber
\end{align}

where $\hat{b}_{\mathbf{m}}^{\dagger}$ creates a phonon of frequency $\omega_{\mathrm{b}}$ at a site ${\bf m} \in (n_x, n_z) \equiv (n_x, 0)$ (with $N_z = 1$) and $\gamma$ is the exciton-phonon coupling strength. We utilize a mixed quantum-classical approach, namely mean-field ehrenfest~\cite{hoffmann2019capturing, ghosh2025mean, li2019comparison, crespo2018recent}, to propagate $\hat{H}_{\mathrm{HTC}}$ (see details in SI). In Fig.~\ref{fig4}g-j we illustrate effect of the exciton-phonon couplings on the coherent frequency conversion via LMME. We find that the inputs and outputs remains phase-locked, similar to the phonon-free case presented in Fig.~\ref{fig4}c-f. Note that this is despite using a relatively strong exciton–phonon coupling ($\gamma = 4\gamma_0$ with $\gamma_0 \approx 7.3067 \times 10^{-5}\,\mathrm{a.u.}$); by comparison, typical polycyclic aromatic hydrocarbons (and their derivatives~\cite{Janke2020JCP}) have $\gamma \approx \gamma_0$. Fig.~\ref{fig4}l illustrates that the relative phases between input and output photons remain the same even in the presence of strong phonon interactions. However, as expected, an increase in exciton–phonon coupling leads to more decoherence marked with increased decay in oscillation in Fig.~\ref{fig4}b. Importantly, decoherence becomes substantial only at extreme  exciton–phonon coupling ($\gamma = 8\gamma_0$). This is expected as we are performing frequency conversion far away from the excitonic transition at $3.2$ eV such that the effective polariton-phonon couplings are substantially suppressed~\cite{chng2025quantum, XuNC2023}.

\section{Conclusion} 
In this work, we introduce and theoretically characterize the light–matter moiré effect (LMME) that emerges when an isotropic 2D material is tilted inside a 3D Fabry–Pérot cavity. Unlike conventional moiré phenomena in twisted multilayer systems, LMME arises purely from geometric modulation of the light–matter coupling. This produces momentum-displaced replicas of the polariton dispersion, in close analogy to the Rashba and Dresselhaus spin-orbit couplings, and enables the formation of anisotropic flat bands near the Brillouin-zone center. 

We demonstrate that LMME can be used to perform coherent frequency conversion, where the relative phase encoded in an initial superposition of two photons is transferred to a superposition of two photons at a different frequency with a frequency shift that is determined by the angle of tilt. Importantly, this process preserves phase information even in the presence of substantial exciton–phonon interactions, highlighting its robust nature and its potential application in quantum information processing.

Beyond its fundamental interest, LMME offers a versatile platform for engineering polariton band structures and tailoring light–matter interactions. The ability to generate tunable flat bands and perform phase-preserving frequency conversion through cavity-material architecture opens new avenues for engineering polariton-based quantum devices, directional transport, and efficient room-temperature polariton condensation.

\section{Acknowledgments}
This work was supported by the Texas A\&M startup funds. This work used TAMU FASTER at the Texas A\&M University through allocation  PHY230021 from the Advanced Cyberinfrastructure Coordination Ecosystem: Services \& Support (ACCESS) program, which is supported by National Science Foundation grants \#2138259, \#2138286, \#2138307, \#2137603, and \#2138296.  The authors appreciate discussions with Ding Xu and Milan Delor.

\bibliography{bib.bib}

\begin{thebibliography}{10}
\expandafter\ifx\csname url\endcsname\relax
  \def\url#1{\texttt{#1}}\fi
\expandafter\ifx\csname urlprefix\endcsname\relax\def\urlprefix{URL }\fi
\providecommand{\bibinfo}[2]{#2}
\providecommand{\eprint}[2][]{\url{#2}}

\bibitem{LiangNP2024}
\bibinfo{author}{Liang, J.} \emph{et~al.}
\newblock \bibinfo{title}{Polariton spin hall effect in a rashba--dresselhaus regime at room temperature}.
\newblock \emph{\bibinfo{journal}{Nature Photonics}} \textbf{\bibinfo{volume}{18}}, \bibinfo{pages}{357--362} (\bibinfo{year}{2024}).

\bibitem{MandalCR2023}
\bibinfo{author}{Mandal, A.} \emph{et~al.}
\newblock \bibinfo{title}{Theoretical advances in polariton chemistry and molecular cavity quantum electrodynamics}.
\newblock \emph{\bibinfo{journal}{Chemical Reviews}} \textbf{\bibinfo{volume}{123}}, \bibinfo{pages}{9786--9879} (\bibinfo{year}{2023}).

\bibitem{XuNC2023}
\bibinfo{author}{Xu, D.} \emph{et~al.}
\newblock \bibinfo{title}{Ultrafast imaging of polariton propagation and interactions}.
\newblock \emph{\bibinfo{journal}{Nature Communications}} \textbf{\bibinfo{volume}{14}}, \bibinfo{pages}{3881} (\bibinfo{year}{2023}).

\bibitem{SanvittoNM2016}
\bibinfo{author}{Sanvitto, D.} \& \bibinfo{author}{K{\'e}na-Cohen, S.}
\newblock \bibinfo{title}{The road towards polaritonic devices}.
\newblock \emph{\bibinfo{journal}{Nature materials}} \textbf{\bibinfo{volume}{15}}, \bibinfo{pages}{1061--1073} (\bibinfo{year}{2016}).

\bibitem{SandikNM2025}
\bibinfo{author}{Sandik, G.}, \bibinfo{author}{Feist, J.}, \bibinfo{author}{Garc{\'\i}a-Vidal, F.~J.} \& \bibinfo{author}{Schwartz, T.}
\newblock \bibinfo{title}{Cavity-enhanced energy transport in molecular systems}.
\newblock \emph{\bibinfo{journal}{Nature Materials}} \textbf{\bibinfo{volume}{24}}, \bibinfo{pages}{344--355} (\bibinfo{year}{2025}).

\bibitem{XiangCR2024}
\bibinfo{author}{Xiang, B.} \& \bibinfo{author}{Xiong, W.}
\newblock \bibinfo{title}{Molecular polaritons for chemistry, photonics and quantum technologies}.
\newblock \emph{\bibinfo{journal}{Chemical Reviews}} \textbf{\bibinfo{volume}{124}}, \bibinfo{pages}{2512--2552} (\bibinfo{year}{2024}).

\bibitem{KockumNRP2019}
\bibinfo{author}{Frisk~Kockum, A.}, \bibinfo{author}{Miranowicz, A.}, \bibinfo{author}{De~Liberato, S.}, \bibinfo{author}{Savasta, S.} \& \bibinfo{author}{Nori, F.}
\newblock \bibinfo{title}{Ultrastrong coupling between light and matter}.
\newblock \emph{\bibinfo{journal}{Nature Reviews Physics}} \textbf{\bibinfo{volume}{1}}, \bibinfo{pages}{19--40} (\bibinfo{year}{2019}).

\bibitem{SunNC2024}
\bibinfo{author}{Sun, K.} \& \bibinfo{author}{Ribeiro, R.~F.}
\newblock \bibinfo{title}{Theoretical formulation of chemical equilibrium under vibrational strong coupling}.
\newblock \emph{\bibinfo{journal}{Nature Communications}} \textbf{\bibinfo{volume}{15}}, \bibinfo{pages}{2405} (\bibinfo{year}{2024}).

\bibitem{NagarajanJACS2021}
\bibinfo{author}{Nagarajan, K.}, \bibinfo{author}{Thomas, A.} \& \bibinfo{author}{Ebbesen, T.~W.}
\newblock \bibinfo{title}{Chemistry under vibrational strong coupling}.
\newblock \emph{\bibinfo{journal}{J. Am. Chem. Soc.}} \textbf{\bibinfo{volume}{143}}, \bibinfo{pages}{16877--16889} (\bibinfo{year}{2021}).

\bibitem{RibeiroCS2018}
\bibinfo{author}{Ribeiro, R.~F.}, \bibinfo{author}{Mart{\'\i}nez-Mart{\'\i}nez, L.~A.}, \bibinfo{author}{Du, M.}, \bibinfo{author}{Campos-Gonzalez-Angulo, J.} \& \bibinfo{author}{Yuen-Zhou, J.}
\newblock \bibinfo{title}{Polariton chemistry: controlling molecular dynamics with optical cavities}.
\newblock \emph{\bibinfo{journal}{Chemical science}} \textbf{\bibinfo{volume}{9}}, \bibinfo{pages}{6325--6339} (\bibinfo{year}{2018}).

\bibitem{BasovNp2021}
\bibinfo{author}{Basov, D.~N.}, \bibinfo{author}{Asenjo-Garcia, A.}, \bibinfo{author}{Schuck, P.~J.}, \bibinfo{author}{Zhu, X.} \& \bibinfo{author}{Rubio, A.}
\newblock \bibinfo{title}{Polariton panorama}.
\newblock \emph{\bibinfo{journal}{Nanophotonics}} \textbf{\bibinfo{volume}{10}}, \bibinfo{pages}{549--577} (\bibinfo{year}{2021}).

\bibitem{LiARPC2022}
\bibinfo{author}{Li, T.~E.}, \bibinfo{author}{Cui, B.}, \bibinfo{author}{Subotnik, J.~E.} \& \bibinfo{author}{Nitzan, A.}
\newblock \bibinfo{title}{Molecular polaritonics: Chemical dynamics under strong light--matter coupling}.
\newblock \emph{\bibinfo{journal}{Annual review of physical chemistry}} \textbf{\bibinfo{volume}{73}}, \bibinfo{pages}{43--71} (\bibinfo{year}{2022}).

\bibitem{ji2025selective}
\bibinfo{author}{Ji, X.} \& \bibinfo{author}{Li, T.~E.}
\newblock \bibinfo{title}{Selective excitation of ir-inactive modes via vibrational polaritons: Insights from atomistic simulations}.
\newblock \emph{\bibinfo{journal}{The Journal of Physical Chemistry Letters}} \textbf{\bibinfo{volume}{16}}, \bibinfo{pages}{5034--5042} (\bibinfo{year}{2025}).

\bibitem{KavokinNRP2022}
\bibinfo{author}{Kavokin, A.} \emph{et~al.}
\newblock \bibinfo{title}{Polariton condensates for classical and quantum computing}.
\newblock \emph{\bibinfo{journal}{Nature Reviews Physics}} \textbf{\bibinfo{volume}{4}}, \bibinfo{pages}{435--451} (\bibinfo{year}{2022}).

\bibitem{KeelingARPC2020}
\bibinfo{author}{Keeling, J.} \& \bibinfo{author}{K\'{e}na-Cohen, S.}
\newblock \bibinfo{title}{Bose–einstein condensation of exciton-polaritons in organic microcavities}.
\newblock \emph{\bibinfo{journal}{Annual Review of Physical Chemistry}} \textbf{\bibinfo{volume}{71}}, \bibinfo{pages}{435--459} (\bibinfo{year}{2020}).

\bibitem{DaskalakisNM2014}
\bibinfo{author}{Daskalakis, K.}, \bibinfo{author}{Maier, S.}, \bibinfo{author}{Murray, R.} \& \bibinfo{author}{K{\'e}na-Cohen, S.}
\newblock \bibinfo{title}{Nonlinear interactions in an organic polariton condensate}.
\newblock \emph{\bibinfo{journal}{Nature materials}} \textbf{\bibinfo{volume}{13}}, \bibinfo{pages}{271--278} (\bibinfo{year}{2014}).

\bibitem{GeorgakilasNC2025}
\bibinfo{author}{Georgakilas, I.} \emph{et~al.}
\newblock \bibinfo{title}{Room-temperature cavity exciton-polariton condensation in perovskite quantum dots}.
\newblock \emph{\bibinfo{journal}{Nature Communications}} \textbf{\bibinfo{volume}{16}}, \bibinfo{pages}{5228} (\bibinfo{year}{2025}).

\bibitem{AlnatahACSP2025}
\bibinfo{author}{Alnatah, H.} \emph{et~al.}
\newblock \bibinfo{title}{Bose–einstein condensation of polaritons at room temperature in a gaas/algaas structure}.
\newblock \emph{\bibinfo{journal}{ACS Photonics}} \textbf{\bibinfo{volume}{12}}, \bibinfo{pages}{48--52} (\bibinfo{year}{2025}).

\bibitem{SeptembrePRL2024}
\bibinfo{author}{Septembre, I.}, \bibinfo{author}{Leblanc, C.}, \bibinfo{author}{Solnyshkov, D.~D.} \& \bibinfo{author}{Malpuech, G.}
\newblock \bibinfo{title}{Topological moir\'{e} polaritons}.
\newblock \emph{\bibinfo{journal}{Phys. Rev. Lett.}} \textbf{\bibinfo{volume}{133}}, \bibinfo{pages}{266602} (\bibinfo{year}{2024}).

\bibitem{SpencerSA2021}
\bibinfo{author}{Spencer, M.~S.} \emph{et~al.}
\newblock \bibinfo{title}{Spin-orbit--coupled exciton-polariton condensates in lead halide perovskites}.
\newblock \emph{\bibinfo{journal}{Science Advances}} \textbf{\bibinfo{volume}{7}}, \bibinfo{pages}{eabj7667} (\bibinfo{year}{2021}).

\bibitem{shi2025coherentOpticalSpinHall}
\bibinfo{author}{Shi, Y.} \emph{et~al.}
\newblock \bibinfo{title}{Coherent optical spin hall transport for polaritonics at room temperature}.
\newblock \emph{\bibinfo{journal}{Nature Materials}} \textbf{\bibinfo{volume}{24}}, \bibinfo{pages}{56--62} (\bibinfo{year}{2025}).

\bibitem{lekenta2018tunableLightSpinHall}
\bibinfo{author}{Lekenta, K.} \emph{et~al.}
\newblock \bibinfo{title}{Tunable optical spin hall effect in a liquid crystal microcavity}.
\newblock \emph{\bibinfo{journal}{Light: Science \& Applications}} \textbf{\bibinfo{volume}{7}}, \bibinfo{pages}{74} (\bibinfo{year}{2018}).

\bibitem{Krupp2025NC}
\bibinfo{author}{Krupp, N.}, \bibinfo{author}{Groenhof, G.} \& \bibinfo{author}{Vendrell, O.}
\newblock \bibinfo{title}{Quantum dynamics simulation of exciton-polariton transport}.
\newblock \emph{\bibinfo{journal}{Nature Communications}} \textbf{\bibinfo{volume}{16}}, \bibinfo{pages}{5431} (\bibinfo{year}{2025}).

\bibitem{ying2025microscopic}
\bibinfo{author}{Ying, W.}, \bibinfo{author}{Chng, B.~X.}, \bibinfo{author}{Delor, M.} \& \bibinfo{author}{Huo, P.}
\newblock \bibinfo{title}{Microscopic theory of polariton group velocity renormalization}.
\newblock \emph{\bibinfo{journal}{Nature Communications}} \textbf{\bibinfo{volume}{16}}, \bibinfo{pages}{6950} (\bibinfo{year}{2025}).

\bibitem{Blackham2025Arxiv}
\bibinfo{author}{Blackham, L.}, \bibinfo{author}{Manjalingal, A.}, \bibinfo{author}{Koshkaki, S.~R.} \& \bibinfo{author}{Mandal, A.}
\newblock \bibinfo{title}{Microscopic theory of polaron-polariton dispersion and propagation}.
\newblock \emph{\bibinfo{journal}{arXiv:2501.16622}}  (\bibinfo{year}{2025}).

\bibitem{YangPNAS2023}
\bibinfo{author}{Yang, Z.}, \bibinfo{author}{Bhakta, H.~H.} \& \bibinfo{author}{Xiong, W.}
\newblock \bibinfo{title}{Enabling multiple intercavity polariton coherences by adding quantum confinement to cavity molecular polaritons}.
\newblock \emph{\bibinfo{journal}{Proceedings of the National Academy of Sciences}} \textbf{\bibinfo{volume}{120}}, \bibinfo{pages}{e2206062120} (\bibinfo{year}{2023}).

\bibitem{CaoN2018}
\bibinfo{author}{Cao, Y.} \emph{et~al.}
\newblock \bibinfo{title}{Correlated insulator behaviour at half-filling in magic-angle graphene superlattices}.
\newblock \emph{\bibinfo{journal}{Nature}} \textbf{\bibinfo{volume}{556}}, \bibinfo{pages}{80--84} (\bibinfo{year}{2018}).

\bibitem{AndreiNRM2021}
\bibinfo{author}{Andrei, E.~Y.} \emph{et~al.}
\newblock \bibinfo{title}{The marvels of moir{\'e} materials}.
\newblock \emph{\bibinfo{journal}{Nature Reviews Materials}} \textbf{\bibinfo{volume}{6}}, \bibinfo{pages}{201--206} (\bibinfo{year}{2021}).

\bibitem{kim2013NM}
\bibinfo{author}{Kim, K.~S.} \emph{et~al.}
\newblock \bibinfo{title}{Coexisting massive and massless dirac fermions in symmetry-broken bilayer graphene}.
\newblock \emph{\bibinfo{journal}{Nature materials}} \textbf{\bibinfo{volume}{12}}, \bibinfo{pages}{887--892} (\bibinfo{year}{2013}).

\bibitem{Xiang2024JCP}
\bibinfo{author}{Xiang, B.} \emph{et~al.}
\newblock \bibinfo{title}{Optical spin hall effect in exciton--polariton condensates in lead halide perovskite microcavities}.
\newblock \emph{\bibinfo{journal}{The Journal of Chemical Physics}} \textbf{\bibinfo{volume}{160}} (\bibinfo{year}{2024}).

\bibitem{MandalNL2023}
\bibinfo{author}{Mandal, A.} \emph{et~al.}
\newblock \bibinfo{title}{Microscopic theory of multimode polariton dispersion in multilayered materials}.
\newblock \emph{\bibinfo{journal}{Nano Letters}} \textbf{\bibinfo{volume}{23}}, \bibinfo{pages}{4082--4089} (\bibinfo{year}{2023}).

\bibitem{sun2025exploring}
\bibinfo{author}{Sun, K.}, \bibinfo{author}{Du, M.} \& \bibinfo{author}{Yuen-Zhou, J.}
\newblock \bibinfo{title}{Exploring the delocalization of dark states in a multimode optical cavity}.
\newblock \emph{\bibinfo{journal}{The Journal of Physical Chemistry C}} \textbf{\bibinfo{volume}{129}}, \bibinfo{pages}{9837--9843} (\bibinfo{year}{2025}).

\bibitem{ghosh2025mean}
\bibinfo{author}{Ghosh, P.}, \bibinfo{author}{Manjalingal, A.}, \bibinfo{author}{Wickramasinghe, S.}, \bibinfo{author}{Koshkaki, S.~R.} \& \bibinfo{author}{Mandal, A.}
\newblock \bibinfo{title}{Mean-field mixed quantum-classical approach for many-body quantum dynamics of exciton-polaritons}.
\newblock \emph{\bibinfo{journal}{arXiv preprint arXiv:2505.04044}}  (\bibinfo{year}{2025}).

\bibitem{zoubi2005PRB}
\bibinfo{author}{Zoubi, H.} \& \bibinfo{author}{La~Rocca, G.~C.}
\newblock \bibinfo{title}{Microscopic theory of anisotropic organic cavity exciton polaritons}.
\newblock \emph{\bibinfo{journal}{Physical Review B—Condensed Matter and Materials Physics}} \textbf{\bibinfo{volume}{71}}, \bibinfo{pages}{235316} (\bibinfo{year}{2005}).

\bibitem{Nguyen2025Arxiv}
\bibinfo{author}{Nguyen, H.}, \bibinfo{author}{Mandal, A.}, \bibinfo{author}{Mahajan, A.} \& \bibinfo{author}{Reichman, D.~R.}
\newblock \bibinfo{title}{Mixed quantum-classical methods for polaron spectral functions}.
\newblock \emph{\bibinfo{journal}{arXiv preprint arXiv:2505.13365}}  (\bibinfo{year}{2025}).

\bibitem{Neves2024NPJCM}
\bibinfo{author}{Neves, P.~M.} \emph{et~al.}
\newblock \bibinfo{title}{Crystal net catalog of model flat band materials}.
\newblock \emph{\bibinfo{journal}{npj Computational Materials}} \textbf{\bibinfo{volume}{10}}, \bibinfo{pages}{39} (\bibinfo{year}{2024}).

\bibitem{cao2018correlated}
\bibinfo{author}{Cao, Y.} \emph{et~al.}
\newblock \bibinfo{title}{Correlated insulator behaviour at half-filling in magic-angle graphene superlattices}.
\newblock \emph{\bibinfo{journal}{Nature}} \textbf{\bibinfo{volume}{556}}, \bibinfo{pages}{80--84} (\bibinfo{year}{2018}).

\bibitem{park2021flavour}
\bibinfo{author}{Park, J.~M.}, \bibinfo{author}{Cao, Y.}, \bibinfo{author}{Watanabe, K.}, \bibinfo{author}{Taniguchi, T.} \& \bibinfo{author}{Jarillo-Herrero, P.}
\newblock \bibinfo{title}{Flavour hund’s coupling, chern gaps and charge diffusivity in moir{\'e} graphene}.
\newblock \emph{\bibinfo{journal}{Nature}} \textbf{\bibinfo{volume}{592}}, \bibinfo{pages}{43--48} (\bibinfo{year}{2021}).

\bibitem{balents2020superconductivity}
\bibinfo{author}{Balents, L.}, \bibinfo{author}{Dean, C.~R.}, \bibinfo{author}{Efetov, D.~K.} \& \bibinfo{author}{Young, A.~F.}
\newblock \bibinfo{title}{Superconductivity and strong correlations in moir{\'e} flat bands}.
\newblock \emph{\bibinfo{journal}{Nature Physics}} \textbf{\bibinfo{volume}{16}}, \bibinfo{pages}{725--733} (\bibinfo{year}{2020}).

\bibitem{koshkaki2025exciton}
\bibinfo{author}{Koshkaki, S.~R.}, \bibinfo{author}{Manjalingal, A.}, \bibinfo{author}{Blackham, L.} \& \bibinfo{author}{Mandal, A.}
\newblock \bibinfo{title}{Exciton-polariton dynamics in multilayered materials}.
\newblock \emph{\bibinfo{journal}{arXiv preprint arXiv:2502.12933}}  (\bibinfo{year}{2025}).

\bibitem{Mann2023qfc}
\bibinfo{author}{Mann, F.}, \bibinfo{author}{Chrzanowski, H.~M.}, \bibinfo{author}{Gewers, F.}, \bibinfo{author}{Placke, M.} \& \bibinfo{author}{Ramelow, S.}
\newblock \bibinfo{title}{Low-noise quantum frequency conversion in a monolithic cavity with bulk periodically poled potassium titanyl phosphate}.
\newblock \emph{\bibinfo{journal}{Phys. Rev. Appl.}} \textbf{\bibinfo{volume}{20}}, \bibinfo{pages}{054010} (\bibinfo{year}{2023}).

\bibitem{wang2023qfc}
\bibinfo{author}{Wang, X.} \emph{et~al.}
\newblock \bibinfo{title}{Quantum frequency conversion and single-photon detection with lithium niobate nanophotonic chips}.
\newblock \emph{\bibinfo{journal}{npj Quantum Information}} \textbf{\bibinfo{volume}{9}}, \bibinfo{pages}{38} (\bibinfo{year}{2023}).

\bibitem{curtz2010coherent}
\bibinfo{author}{Curtz, N.}, \bibinfo{author}{Thew, R.}, \bibinfo{author}{Simon, C.}, \bibinfo{author}{Gisin, N.} \& \bibinfo{author}{Zbinden, H.}
\newblock \bibinfo{title}{Coherent frequency-down-conversion interface for quantum repeaters}.
\newblock \emph{\bibinfo{journal}{Optics Express}} \textbf{\bibinfo{volume}{18}}, \bibinfo{pages}{22099--22104} (\bibinfo{year}{2010}).

\bibitem{Abdo2013CFC}
\bibinfo{author}{Abdo, B.} \emph{et~al.}
\newblock \bibinfo{title}{Full coherent frequency conversion between two propagating microwave modes}.
\newblock \emph{\bibinfo{journal}{Phys. Rev. Lett.}} \textbf{\bibinfo{volume}{110}}, \bibinfo{pages}{173902} (\bibinfo{year}{2013}).

\bibitem{hoffmann2019capturing}
\bibinfo{author}{Hoffmann, N.~M.}, \bibinfo{author}{Sch{\"a}fer, C.}, \bibinfo{author}{Rubio, A.}, \bibinfo{author}{Kelly, A.} \& \bibinfo{author}{Appel, H.}
\newblock \bibinfo{title}{Capturing vacuum fluctuations and photon correlations in cavity quantum electrodynamics with multitrajectory ehrenfest dynamics}.
\newblock \emph{\bibinfo{journal}{Physical Review A}} \textbf{\bibinfo{volume}{99}}, \bibinfo{pages}{063819} (\bibinfo{year}{2019}).

\bibitem{li2019comparison}
\bibinfo{author}{Li, T.~E.}, \bibinfo{author}{Chen, H.-T.} \& \bibinfo{author}{Subotnik, J.~E.}
\newblock \bibinfo{title}{Comparison of different classical, semiclassical, and quantum treatments of light--matter interactions: Understanding energy conservation}.
\newblock \emph{\bibinfo{journal}{Journal of Chemical Theory and Computation}} \textbf{\bibinfo{volume}{15}}, \bibinfo{pages}{1957--1973} (\bibinfo{year}{2019}).

\bibitem{crespo2018recent}
\bibinfo{author}{Crespo-Otero, R.} \& \bibinfo{author}{Barbatti, M.}
\newblock \bibinfo{title}{Recent advances and perspectives on nonadiabatic mixed quantum--classical dynamics}.
\newblock \emph{\bibinfo{journal}{Chemical reviews}} \textbf{\bibinfo{volume}{118}}, \bibinfo{pages}{7026--7068} (\bibinfo{year}{2018}).

\bibitem{Janke2020JCP}
\bibinfo{author}{Janke, S.~M.}, \bibinfo{author}{Qarai, M.~B.}, \bibinfo{author}{Blum, V.} \& \bibinfo{author}{Spano, F.~C.}
\newblock \bibinfo{title}{Frenkel--holstein hamiltonian applied to absorption spectra of quaterthiophene-based 2d hybrid organic--inorganic perovskites}.
\newblock \emph{\bibinfo{journal}{The Journal of Chemical Physics}} \textbf{\bibinfo{volume}{152}} (\bibinfo{year}{2020}).

\bibitem{chng2025quantum}
\bibinfo{author}{Chng, B.~X.}, \bibinfo{author}{Mondal, M.~E.}, \bibinfo{author}{Ying, W.} \& \bibinfo{author}{Huo, P.}
\newblock \bibinfo{title}{Quantum dynamics simulations of exciton polariton transport}.
\newblock \emph{\bibinfo{journal}{Nano Letters}} \textbf{\bibinfo{volume}{25}}, \bibinfo{pages}{1617--1622} (\bibinfo{year}{2025}).

\end{thebibliography}
\bibliographystyle{naturemag}

\end{document}